\documentclass[prd,twocolumn,aps,noshowpacs,nofootinbib,amsmath,amssymb,floatfix,superscriptaddress]{revtex4}
\usepackage[colorlinks=true,linkcolor=red,citecolor=blue]{hyperref}
\usepackage{amsmath}
\usepackage{amsfonts}
\usepackage{graphicx}
\usepackage{subfigure}
\usepackage{dcolumn}
\usepackage{bm}
\usepackage{booktabs}
\usepackage[utf8]{inputenc}
\usepackage{multirow}
\usepackage{graphicx,graphics,dcolumn,booktabs,bm}
\usepackage{longtable,lscape}
\usepackage{txfonts}
\usepackage{overpic}
\usepackage{amssymb}
\usepackage{indentfirst}
\usepackage{epsfig}
\usepackage{feynmf}   
\usepackage{epstopdf}   
\usepackage{slashed}  
\usepackage{color}
\usepackage[section]{placeins}

\begin{document}

\title{ Next-to-leading order QCD calculation of $B_c$ to charmonium  tensor form factors }
\author{ Wei Tao\footnote{taowei@njnu.edu.cn}}
\author{Zhen-Jun Xiao \footnote{Corresponding author: xiaozhenjun@njnu.edu.cn} }
\author{ Ruilin Zhu~\footnote{Corresponding author: rlzhu@njnu.edu.cn} }
\affiliation{ Department of Physics and Institute of Theoretical Physics, Nanjing Normal University, Nanjing, Jiangsu 210023, China}
\date{\today}
\vspace{0.3in}

\begin{abstract}
We present a next-to-leading order (NLO) QCD corrections to $B_c\to \eta_c$ and $B_c\to J/\psi$ tensor form factors within nonrelativistic QCD (NRQCD) framework.
The full analytical results for $B_c$ to S-wave charmonium  tensor form factors are obtained. We also studied the asymptotic behaviours of tensor form factors in hierarchy heavy quark limit, i.e. $m_b\to\infty,~ m_c\to\infty, ~\mathrm{and }~m_c/m_b\to0$. A compact  expression for tensor form factors  are given analytically in the hierarchy heavy quark limit. The relation among different form factors
 is also analyzed especially at large momentum recoil point. The numerical results for the $B_c$ to charmonium  tensor form factors in all the physical region are given in the end.
 \vspace{0.3in}

\begin{description}
\item[PACS numbers] 12.38.Bx,  13.25.Gv, 14.40.Pq
\end{description}
\end{abstract}

\maketitle

\section{Introduction}
Testing the Standard Model and hunting for new physics is a primary task in particle physics. In recent years, the $b\to c$ transition has
been employed as a vivid window to indirectly detect the possible pattern of new physics. Particularly the $R(D^{(*)})$ and $R(J/\psi)$ anomalies
in recent flavor physics experiments
shall challenge the lepton universality and indicate the possible pattern of new physics~\cite{BaBar:2012obs,Belle:2016dyj,LHCb:2017rln,LHCb:2017vlu}. To distinguish new physics signal from background in these heavy flavor quark decay channels, a precision calculation and analysis of transition form factors is required~\cite{Bernlochner:2021vlv,Harrison:2020nrv,Cheung:2020sbq,Huang:2018nnq}.

The $b\to c$ transition modes in the $B_c$ meson has been studied in lots of frameworks: the lattice QCD simulations~\cite{Colquhoun:2016osw,Harrison:2020gvo}, the nonrelativistic QCD(NRQCD) approach~\cite{Chang:1992pt,Kiselev:2001zb,Bell:2005gw,Qiao:2011yz,Qiao:2012vt,Qiao:2012hp,Zhu:2017lqu,Zhu:2017lwi}, the perturbative QCD approach~\cite{Du:1988ws,Sun:2008ew,Wen-Fei:2013uea,Rui:2014tpa,Liu:2020upy}, the principle of maximum conformality~\cite{Shen:2014msa}, the QCD sum rules~\cite{Colangelo:1992cx,Kiselev:1999sc,Azizi:2009ny}, the light-cone sum rules~\cite{Huang:2007kb},  the light-front quark model~\cite{Wang:2008xt,Ke:2013yka}, the relativistic quark model~\cite{Nobes:2000pm,Ebert:2003cn,Ivanov:2005fd,Ebert:2010zu,Nayak:2022gdo}, the nonrelativistic constituent quark model~\cite{Hernandez:2006gt} and the SU(3) symmetry~\cite{Zhu:2018epc,He:2016xvd}. It is a remarkable progress that the HPQCD collaboration have gained the first lattice QCD results for the $B_c\to J/\psi$ vector
and axial-vector form factors in the full $q^2$ range~\cite{Harrison:2020gvo}. Using the lattice QCD computation of the  $B_c\to J/\psi$  form factors, the HPQCD collaboration then determine
the standard model predictions of $R(J/\psi)$ and improve the theoretical precision. Therein the lattice QCD
results have reduced the tension of $R(J/\psi)$ anomalies and also indicated the LHCb data has a $1.8\sigma$ deviation from the standard model prediction. To include the possible new physics, other form factors such as scalar, pseudoscalar, and tensor form factors  are also involved in the processes, apart from vector and axial-vector form factors.
These new form factors are not simulated in lattice QCD currently. Fortunately, one can perturbatively calculate these form factors at large momentum recoil order by order in NRQCD approach.

NRQCD is a powerful theoretical framework to deal with the production and decay of double heavy quark system~\cite{Bodwin:1994jh}. There are three kinds of typical scales ordered by
the quark relative velocity $v$: the heavy quark mass ($m_Q$), around and above which the perturbative interactions dominate for the hadron production and decay;
the heavy quark relative momentum ($m_Q v$); the heavy quark kinetic energy ($m_Q v^2$), around which the nonperturbative binding dominate.
The form factors can be expressed by the series of nonperturbative long-distance matrix elements (LDMEs) and the corresponding perturbative Wilson coefficients. In this paper,
 we major focus on the next-to-leading order (NLO) QCD corrections to the $B_c\to \eta_c$ and $B_c\to J/\psi$ form factors. The scalar and pseudoscalar form factors can
 be obtained from vector and axial-vector form factors by equation of motion.  Thus we will calculate the  $B_c\to \eta_c$ and $B_c\to J/\psi$ tensor form factors at NLO in NRQCD framework.

 Even though the definition of form factors  only relies on the local bilinear current, there needs a new renormalization factor to cancel the UV divergence since the tensor current
 is not a conserved  current. We will check the UV and IR behaviour  of the  $B_c\to \eta_c$ and $B_c\to J/\psi$  tensor form factors. On the other hand, we will investigate the
 relation among different form factors in the hierarchy heavy quark limit. Previous studies have indicate that there are degenerate for vector and axial-vector form factors in hierarchy heavy quark limit. Very similarly to Isgur-Wise function in small momentum recoil, the form factors are  not independent at large momentum recoil. Thus we will  check
 the asymptotic expressions of tensor form factors in the hierarchy heavy quark limit.

The paper is arranged as follows. We give the definition and the LO results for $B_c\to \eta_c$ and $B_c\to J/\psi$  tensor form factors in Section II.
We present the NLO QCD corrections to the  $B_c\to \eta_c$ and $B_c\to J/\psi$  tensor form factors, discuss the UV and IR behaviours, and
give the asymptotic analysis  of tensor form factors in the hierarchy heavy quark limit  in Section III. Numerical results and discussions are given in  Section IV.
In the end we give the conclusion.

\section{\texorpdfstring{$B_c\to \eta_c$}{} and \texorpdfstring{$B_c\to J/\psi$}{} tensor form factors }

Inputting various Dirac Gamma matrixes in  bilinear local quark current sandwiched between the $B_c$ meson and a charmonium states, one can define various
form factors. The tensor form factors for  $B_c$ meson into a S-wave charmonium are defined as~\cite{Isgur:1990kf,Ball:1998kk,Ali:1999mm,Beneke:2000wa}

\begin{align}\label{ff1}
&\langle \eta_{c}(p)\vert  \bar c \sigma^{\mu\nu}q_\nu b \vert B_{c}(P)\rangle\nonumber\\&
=\frac{f_{T}(q^{2})}{
m_{B_{c}}+m_{\eta_{c}}}\left(q^{2}(P^{\mu}+p^{\mu})-(m_{B_{c}}^{2}-
m_{\eta_{c}}^{2})q^{\mu}\right)\,,
\\&
\langle J/\psi(p,\varepsilon^{*})\vert \bar c \sigma^{\mu\nu}q_\nu b\vert
B_{c}(P)\rangle=
2i T_1(q^{2})\epsilon^{\mu\nu\rho\sigma}
\varepsilon_{\nu}^{*}p_{\rho}P_{\sigma}\,,\label{eq2}
\\&
\langle J/\psi(p,\varepsilon^{*})\vert \bar c \sigma^{\mu\nu}\gamma^5 q_\nu   b\vert B_{c}(P)\rangle \nonumber\\& =T_2(q^{2})\left((m_{B_{c}}^{2}-m_{J/\psi}^{2})\varepsilon^{*\mu}-\varepsilon^{*}\cdot q(P^{\mu}+p^{\mu})\right)
\nonumber\\&
~~~~+T_{3}(q^{2})\varepsilon^{*}\cdot
q\left( q^{\mu}-\frac{q^{2}}{m_{B_{c}}^{2}-m_{J/\psi}^{2}}(P^{\mu}+p^{\mu})\right)\label{eq3}
\,,
\end{align}
where Dirac operator $\sigma^{\mu\nu}=\frac{i}{2}(\gamma^\mu \gamma^\nu-\gamma^\nu \gamma^\mu)$.
We denote the momentum transfer as $q=P-p$ and we have the physical constraint $0\leq q^2\leq (m_{B_c}- m_{J/\psi(\eta_c)})^2$ in form factors.
The $m$ and $\varepsilon$ are the mass and polarization vector of the mesons. We also use the convention of Levi-Civita tensor $\epsilon^{0123}=1$.
Note  that  $T_1(0)=T_2(0)$  by using the identities
$
\sigma_{\mu\nu}\gamma_5 = \frac{i}{2}\,\epsilon_{\mu\nu\rho\sigma}
\,\sigma^{\rho\sigma}
$
and
$
\epsilon_{\mu\nu\rho\sigma}\,\sigma^{\mu\nu}\gamma_5 = -2 i\,\sigma_{\rho\sigma}
$ in Eqs.~(\ref{eq2}) and (\ref{eq3}).

\begin{figure}[htbp]
\begin{center}
\includegraphics[width=0.95\linewidth]{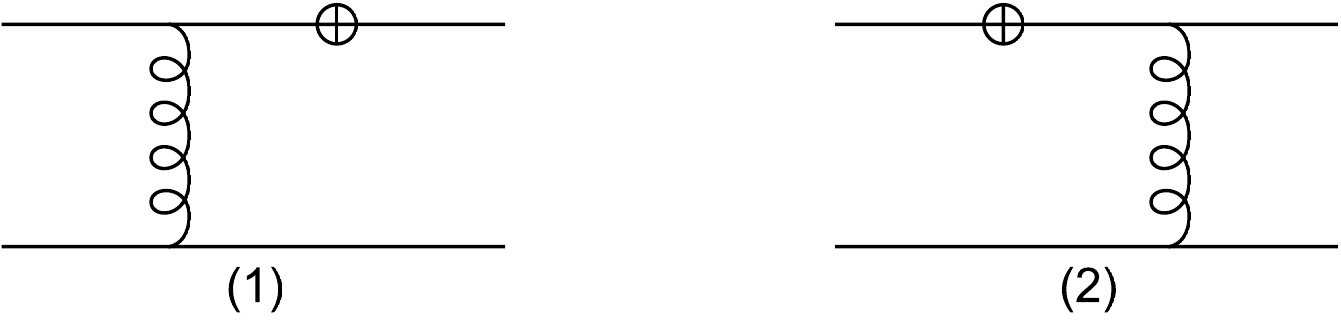}
\caption{\label{fig:bcpicstree}Tree level diagrams for the form factors of $B_c$ into a $S$-wave charmonium, where the symbol   ``$\oplus$'' denotes certain current operators and
the lower line is the spectator charm quark. At LO, one gluon is exchanged between the upper bottom/charm quark and the lower charm quark.}
\end{center}
\end{figure}

In NRQCD, both the $B_c$ meson and $J/\psi$ can be treated as nonrelativistic bound states. The decay amplitudes for $B_c\to J/\psi$ can be factorized as the short-distance Wilson coefficients and the LDMEs~\cite{Bodwin:1994jh,Chang:1992pt,Bell:2006tz,Qiao:2012hp}. The LO Feynman diagrams are plotted in Fig.~\ref{fig:bcpicstree}. Using NRQCD, the leading order results for the form factors are
\begin{widetext}
\begin{eqnarray}
f^\text{LO}_{T}(z,s)
&=&\frac{16 \sqrt{2} C_A  C_F  \pi  s^2 (z+1)^{3/2} (3
   z+1) \alpha _s\psi(0)_{B_c}
\psi(0)_{\eta_c}}{z^{3/2}\left(s z^2-2 s z+1\right)^2 m_b^3 N_c }\,,\\
T^\text{LO}_{1}(z,s)
&=&\frac{4 \sqrt{2}  C_A  C_F \pi  s \sqrt{z+1}\left(5 s z^2+6 s z+4
   s+1\right) \alpha _s \psi(0)_{B_c}\psi(0)_{J/\psi}}{z^{3/2}\left(s z^2-2 s z+1\right)^2 m_b^3 N_c }\,,\\
T^\text{LO}_{2}(z,s)
&=&\frac{4 \sqrt{2}  C_A  C_F\pi  \sqrt{z+1}
 \left(15 s^2 z^4+8 s^2 z^3-8 s^2 z^2-16 s^2 z+6 s z^2-4 s-1\right) \alpha _s \psi(0)_{B_c}\psi(0)_{J/\psi}}{(z-1) z^{3/2} (3 z+1) \left(s
   z^2-2 s z+1\right)^2 m_b^3 N_c}\,,\\
T^\text{LO}_{3}(z,s)
&=&-\frac{4 \sqrt{2} C_A  C_F  \pi  s \sqrt{z+1}
   \left(3 s z^2+2 s z-4 s-1\right) \alpha _s\psi(0)_{B_c}\psi(0)_{J/\psi}}{z^{3/2}\left(s z^2-2 s z+1\right)^2 m_b^3 N_c }\,,
\end{eqnarray}
\end{widetext}
where $z=m_c/m_b$ and $s=1/(1-q^2/m_b^2)$.
The nonperturbative parameters $\psi(0)_{B_c}$ and $\psi(0)_{J/\psi(\eta_c)}$ are the Schr\"{o}dinger wave functions at
the origin for $b\bar{c}$ and $c\bar{c}$ systems, respectively, which are related to the NRQCD LDMEs for the production and decay processes~\cite{Bodwin:1994jh}.

It is noted that the heavy quark symmetry is involved at leading power in heavy quark effective theory and the form factors at minimum momentum recoil point can be expressed by
the Isgur-Wise functions. It indicates that the heavy-to-heavy transition form factors are not independent in heavy quark symmetry. In this paper, we will  calculate perturbatively the form factors of $B_c$ into a $S$-wave charmonium. The perturbative calculation results are thought to be solid at the maximum
momentum recoil region. Then we can also investigate the asymptotic behaviors in hierarchy heavy  quark limit. We introduce the hierarchy heavy  quark limit, i.e. $m_b\to\infty,~ m_c\to\infty, ~\mathrm{and }~z=m_c/m_b\to 0$ to observe the asymptotic behaviors. One can assume the heavy  quark mass approaching the infinity as $m_c=x^a|_{x\to\infty,a>0}$ and $m_b=x^b|_{x\to\infty,a>0}$, and then $z=m_c/m_b=x^{a-b}|_{x\to\infty,a<b}\to 0$. One can easily
see that the form factors are not independent in this limit. In the following we list the asymptotic
expression for the LO tensor form factors
\begin{eqnarray}
f^\text{Asymp. LO}_{T}(z,s)&=&\frac{16 \sqrt{2} C_A  C_F  \pi  s^2  \alpha _s\psi(0)_{B_c}\psi(0)_{\eta_c}}{z^{3/2} m_b^3 N_c }\,,\\
T^\text{Asymp. LO}_{1}(z,s)&=&\frac{4 \sqrt{2}  C_A  C_F \pi  s \left(4
   s+1\right) \alpha _s \psi(0)_{B_c}\psi(0)_{J/\psi}}{z^{3/2} m_b^3 N_c }\,.\nonumber\\
\end{eqnarray}

\begin{figure*}[htbp]     
\begin{center}
\includegraphics[width=0.85\linewidth]{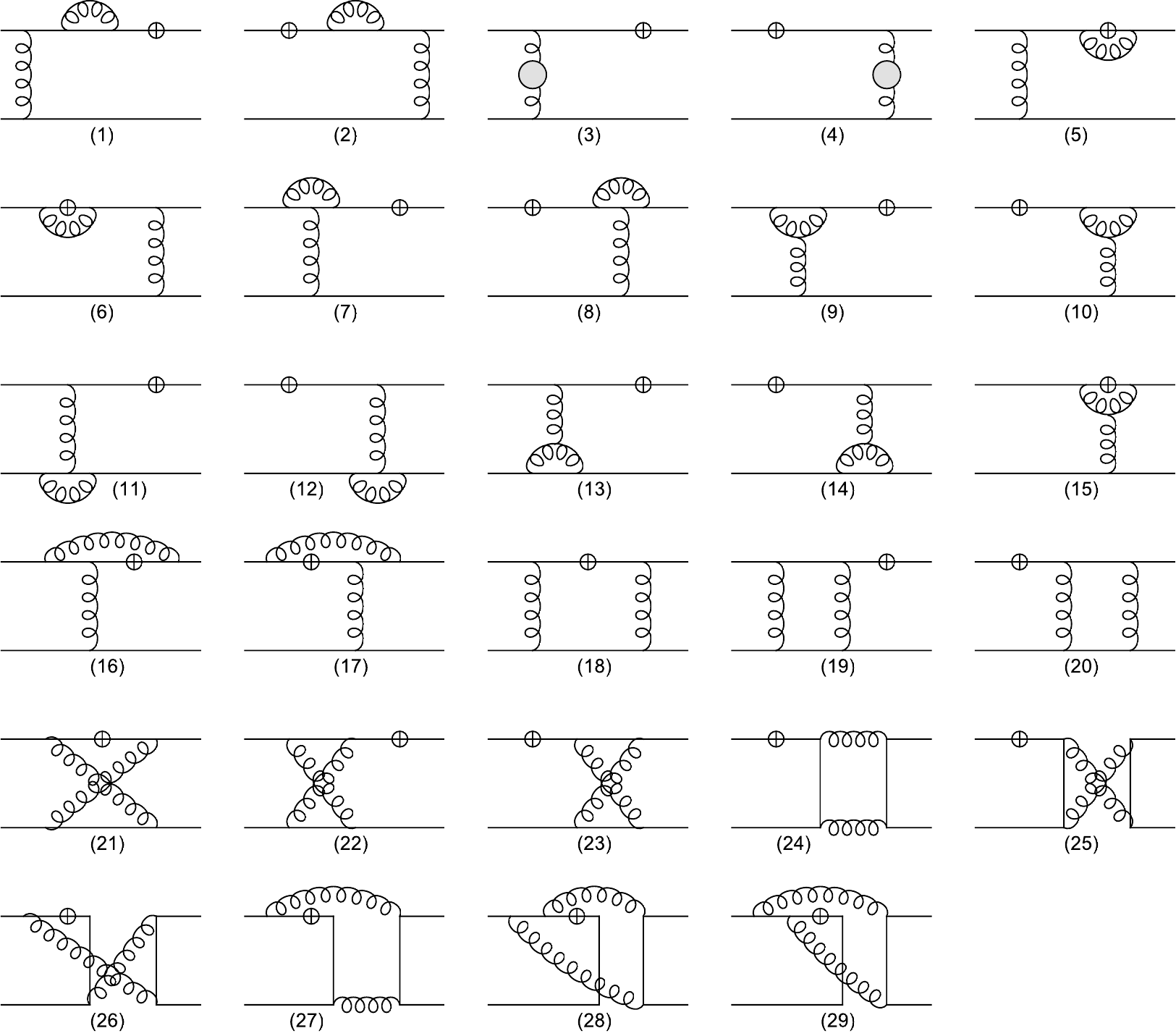}
\caption{\label{fig:bcpics1loop} All of 1-loop  diagrams for the form factors of $B_c \to J/\psi(\eta_c)$, where the symbol ``$\oplus$'' denotes certain current operators.
The bubble in the sub-diagrams (3-4) represents the 1-loop gluon self-energies. The sub-diagrams (24-29) only contribute to $B_c \to \eta_c$ channel.}
\end{center}
\end{figure*}

The tensor form factors $T_2$ and $T_3$ are related to $T_1$ as
\begin{eqnarray}
T^\text{Asymp. LO}_{2}(z,s)
&=&\frac{T^\text{Asymp. LO}_{1}(z,s)}{ s}\,,\\
T^\text{Asymp. LO}_{3}(z,s)
&=&T^\text{Asymp. LO}_{1}(z,s)\,.
\end{eqnarray}

The higher order QCD and relativistic corrections for the  vector and axial-vector form factors for  $B_c$ meson into a S-wave charmonium  can be found in Refs.~\cite{Bell:2006tz,Qiao:2011yz,Qiao:2012vt,Zhu:2017lqu}. We have confirmed all the previous results for the  NLO corrections to the vector and axial-vector form factors. The relativistic corrections for the tensor form factors for  $B_c\to J/\psi$ have been performed in Ref.~\cite{Shen:2021dat}.
In the following section we will study the NLO QCD corrections to
the tensor form factors for both $B_c\to \eta_c$ and $B_c\to J/\psi$ channels. The precision predictions of various form factors shall improve the standard model theoretical uncertainty
and determine the possible pattern of new physics in $R(\eta_c)$ and $R(J/\psi)$ observables.

\section{QCD correction to \texorpdfstring{$B_c\to \eta_c, J/\psi$}{}  tensor Form factors }

We next calculate the NLO QCD corrections to the tensor form factors of $B_c\to (\eta_c, J/\psi)$ transitions.
At LO, the form factors come from two tree diagrams in Fig.~\ref{fig:bcpicstree}.
At NLO, the form factors receive contributions from various 1-loop Feynman diagrams in Fig.~\ref{fig:bcpics1loop}. These  1-loop diagrams include the self-energy
correction, vertex correction, box and pentagon correction.

On the calculation of the 1-loop diagrams, we adopt the Feynman gauge and use dimensional regularization to regularize the  occurring  UV  and  IR divergences.
First, we apply the package FeynArts \cite{Hahn:2000kx} to generate the corresponding Feynman diagrams and amplitudes. We implement the package
FeynCalc \cite{Shtabovenko:2020gxv} to handle amplitudes, i.e., contract indexes, simplify Dirac Gamma matrixes, and obtain  traces.
Then employing the package Apart~\cite{Feng:2012iq} for partial fractions, the full 1-loop amplitudes, including the self-correction,
vertex correction, box and pentagon correction are expressed as the linear combination of the standard Passarino-Veltman scalar integrals A0, B0, C0,
D0 \footnote{However, the five-point integrals in the  sub-diagram (18)  in Fig.~\ref{fig:bcpics1loop}, can only be reduced to A0,B0,C0,D0 by integration
by Parts(IBP) \cite{Chetyrkin:1981qh} without setting scaleless integrals to zero to distinguish UV and IR divergences.}.  We use Package-X \cite{Patel:2016fam} to analytically calculate these Feynman integrals.

The 1-loop self-energy and vertex correction diagrams have the UV divergences, which are thought to be cancelled by the counter-term in standard high-order calculation procedure. But
an additional renormalization factor $Z_{\Gamma}$ for certain current is also required.  The renormalization
constants include $Z_{2}$, $Z_{3}$, $Z_{m}$, $Z_{g}$, and $Z_{\Gamma}$ (see Ref.~\cite{Bauer:2000yr,Bell:2006tz}), referring
to quark field, gluon field, quark mass, strong coupling constant $g_{s}$,  and tensor current respectively. In our calculation the $Z_{3}$, $Z_{g}$, $Z_{\Gamma}$
are defined in the modified-minimal-subtraction ($\mathrm{\overline{MS}}$) scheme, while for $Z_{2}$ and $Z_{m}$ the
on-shell ($\mathrm{OS}$) scheme is employed, which tells
\begin{eqnarray}
&&\hspace{-0.3cm}\delta Z_m=-3C_F
\frac{\alpha_s}{4\pi}\left[\frac{1}{\epsilon_{UV}}+
\ln\frac{\mu^{2}}{m^{2}}+\frac{4}{3}+{\mathcal{O}}(\epsilon)\right]+{\mathcal{O}}(\alpha^2 _s)\;
,
\\ &&\hspace{-0.3cm}\delta Z_2=-C_F
\frac{\alpha_s}{4\pi}\left[\frac{1}
{\epsilon_{UV}}+\frac{2}{\epsilon_{IR}}+3\ln
\frac{\mu^{2}}{m^{2}}+4+{\mathcal{O}}(\epsilon)\right]+{\mathcal{O}}(\alpha^2 _s)\; ,\nonumber\\\\
&&\hspace{-0.3cm}\delta Z_3= \frac{\alpha_s}{4\pi}
\left[(\beta_0-2C_A)\frac{1}{\epsilon_{UV}}+{\mathcal{O}}(\epsilon)\right]+{\mathcal{O}}(\alpha^2 _s)\; ,\\
&&\hspace{-0.3cm}\delta Z_g=-\frac{\beta_0}{2}
\frac{\alpha_s}{4\pi}\left[\frac{1}{\epsilon_{UV}}+{\mathcal{O}}(\epsilon)\right]+{\mathcal{O}}(\alpha^2 _s)\; ,\\
&&\hspace{-0.3cm}\delta Z_{\Gamma}=C_F
\frac{\alpha_s}{4\pi}\left[\frac{1}
{\epsilon_{UV}}+{\mathcal{O}}(\epsilon)\right]+{\mathcal{O}}(\alpha^2 _s)\; .
\end{eqnarray}
Here, $\delta Z_i=Z_i-1$. $\beta_{0}=(11/3)C_{A}-(2/3)n_{f}$ is the one-loop
coefficient of the QCD beta function, $\mu$ is the renormalization scale, and note that $\delta Z_{\Gamma}$ will vanish for vector and axial-vector currents.

It is noted that these renormalization constants are involved in certain counter-term diagrams, but some of them may disappear in the final renormalization formulae since
they will cancel by each other. Now we can write down the renormalization formula for the form factors.
Take the $B_c\to \eta_c$ transition matrix element for example:
\begin{eqnarray}\label{A1}
\langle \eta_{c}\vert  \bar c \Gamma b \vert B_{c}\rangle& =&
(-ig_s)^2\int\int  \mbox{d}^4 x  \mbox{d}^4 y \nonumber\\&&
\langle \eta_{c}|\mbox{T}A^{\mu}(x)A^{\nu}(y) j_\mu(x)j_\nu(y)\, (\bar c \Gamma b) (0)| B_c\rangle\nonumber\\
& =&Z_g^2(-ig^R_s)^2\int\int  \mbox{d}^4 x  \mbox{d}^4 y \int\frac{ \mbox{d}^4 k}{(2\pi)^4}\frac{ e^{-ik\cdot (x-y)}}{k^2+i0}\nonumber\\&&
Z_\Gamma Z_{2,c}^{3/2}Z_{2,b}^{1/2}\langle \eta_{c}|\mbox{T} j^R_\mu(x)j^{R,\mu}(y)\, (\bar c \Gamma b)^R (0)| B_c\rangle^R\,,\nonumber\\
\end{eqnarray}
where the renormalized matrix element has been labelled by a sub-letter $R$. The heavy quark mass which is not explicitly written out should be also renormalized. $j_\mu$ is the conserved heavy quark vector current which does not need the
renormalization, i.e. $j_\mu=j^R_\mu$ for conserved current. Similarly, $Z_\Gamma=1$ for the flavor-changed vector and axial-vector current.
While we have $Z_\Gamma=1+\delta Z_{\Gamma}\neq 1$ for the flavor-changed tensor current.

\begin{figure}
\centering
\includegraphics[width=0.45\textwidth]{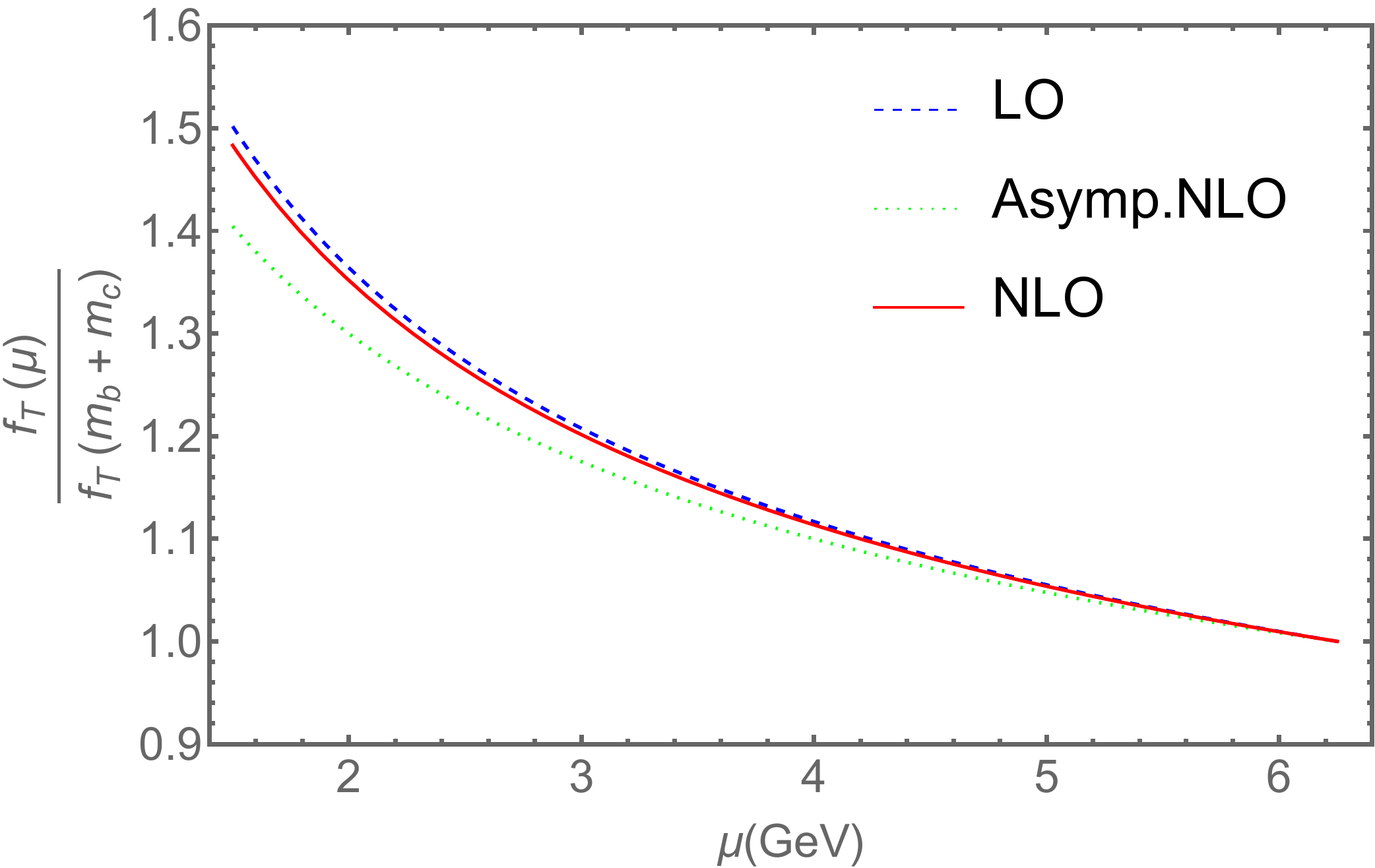}
\caption{The renormalization scale dependence of the form factor $f_T$ at LO, asymptotic NLO and complete  NLO. We set the form
 factors at the maximum recoil point, $q^2=0$. Herein $\mu$ runs from $m_c$ to $m_b+m_c$ with fixed quark mass $m_c=1.5\mathrm{GeV}$ and $ m_b=4.75\mathrm{GeV}$.}
\label{fig:mudepend1}
\vspace{-0mm}
\end{figure}

\begin{figure}
\centering
\includegraphics[width=0.45\textwidth]{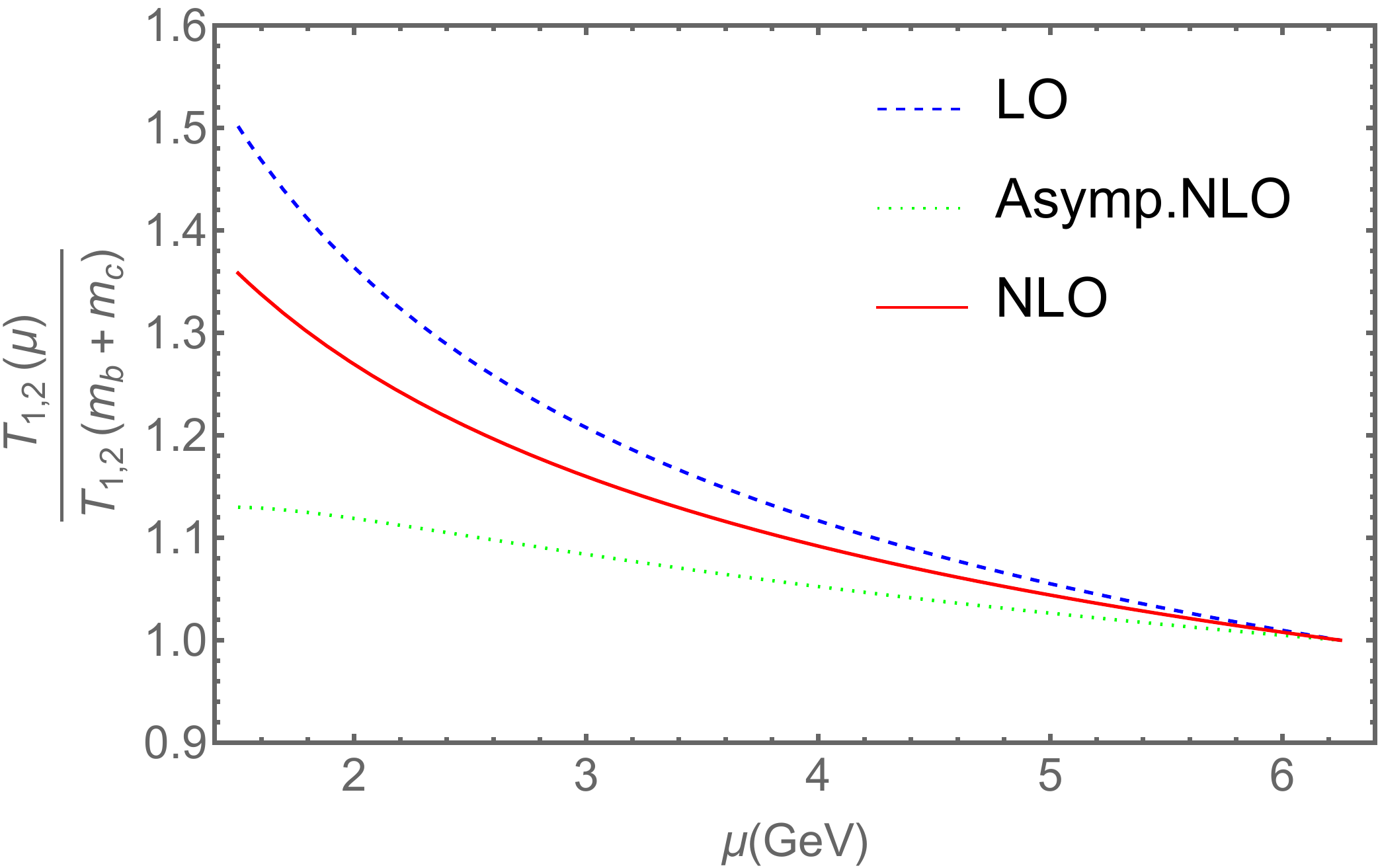}
\caption{The renormalization scale dependence of the form factors $T_1$ and $T_2$ at LO, asymptotic NLO and complete  NLO. We set the form
 factors at the maximum recoil point, $q^2=0$,  which leads to $T_1(q^2=0)=T_2(q^2=0)$. Herein $\mu$ runs from $m_c$ to $m_b+m_c$ with fixed quark mass $m_c=1.5\mathrm{GeV}$ and $ m_b=4.75\mathrm{GeV}$.}
\label{fig:mudepend2}
\vspace{-0mm}
\end{figure}

 After summing up all of the contributions, we find both the UV and IR poles indeed cancel respectively, and obtain complete analytical finite results of the form factors.
 At last, we use Mathematica Function Series to obtain asymptotic  expressions  of the form factors in the hierarchy heavy  quark limit.
 All of the analytical calculations have been numerically checked by the Package AMFlow~\cite{Liu:2022chg} and Package FIESTA~\cite{Smirnov:2021rhf},
 which are consistent with each other.

 The asymptotic  expressions  of form factors in the hierarchy heavy  quark limit are presented in the Appendix. Note that the NLO QCD correction to $B_c\to \eta_c$ tensor form factor $f_T$
has been investigated in Ref.~\cite{Bell:2006tz}. We have confirmed their results of the form factors in the paper~\cite{Bell:2006tz}.
In addition,  we also obtained the NLO QCD correction to $B_c\to J/\psi$ tensor form factors $T_{1,2,3}$, which are important input to precisely study
the $R(J/\psi)$ anomaly~\cite{Tang:2022nqm}. The method may also apply in the transition of double heavy diquark system~\cite{Qin:2021wyh}.

\begin{figure}[thb]    
\centering
\includegraphics[width=0.45\textwidth]{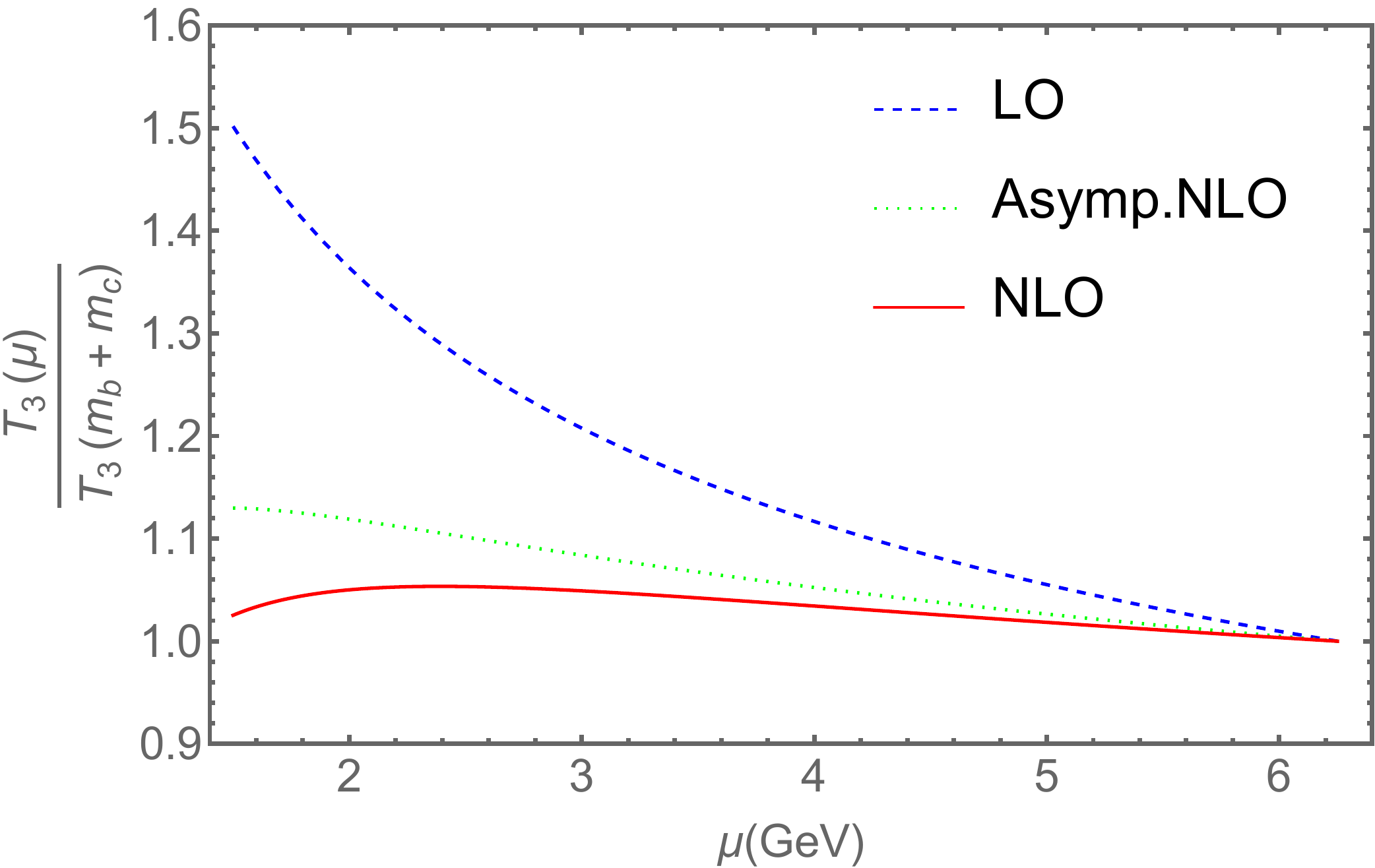}
\caption{The renormalization scale dependence of the form factors $T_3$ at LO, asymptotic NLO and complete  NLO. We set the form
 factors at the maximum recoil point, $q^2=0$. Herein $\mu$ runs from $m_c$ to $m_b+m_c$ with fixed quark mass $m_c=1.5\mathrm{GeV}$ and $ m_b=4.75\mathrm{GeV}$.}
\label{fig:mudepend3}
\end{figure}

\section{ Numerical results and discussions }

In the following numerical calculation, the one loop result for strong coupling constant is used, i.e.
\begin{eqnarray}
\alpha_s(\mu)=\frac{4\pi}{(\frac{11}{3}C_A-\frac{2}{3}n_f)\ln(\frac{\mu^2}{\Lambda_{QCD}^2})}\, ,
\end{eqnarray}
where  the typical QCD scale $\Lambda_{QCD}$  is related to $n_f$. For example $\Lambda^{n_f=5}_{QCD}=87MeV$ is determined by $\alpha_s(m_Z)=0.1179$ with $m_Z=91.1876\mathrm{GeV}$.
 $\Lambda_{QCD}$ will increase if one use the high-order
result for strong coupling constant, however, it is not necessary and only required if we also take the high-order corrections to the form factors. Because we have treated the $B_c$ meson and the
S-wave charmonium as nonrelativistic bound states,  the pole mass of heavy flavor quarks is adopted as: $m_b=4.75\pm 0.05\mathrm{GeV}$ and $m_c=1.5\pm 0.05\mathrm{GeV}$.

First we investigate the renormalization scale dependence of the form factors. To eliminate the uncertainty of nonperturbative NRQCD LDMEs, we define
$f_T(\mu)/f_T(m_b+m_c)$ and $T_i(\mu)/T_i(m_b+m_c)$ which are independent on the nonperturbative NRQCD LDMEs.  We then plot the renormalization
scale dependence of tensor form factors at the LO, asymptotic NLO and complete  NLO results in Figs.~\ref{fig:mudepend1}, \ref{fig:mudepend2}, and \ref{fig:mudepend3}.
In general, the scale dependence at NLO is obviously depressed relative to the LO case.
Naming, the $ \beta_0\alpha_s^2\ln(\mu^2)$ terms in the form factors are cancelled by the  scale dependence in the strong coupling constant. But an additional renormalization
constant $Z_\Gamma$ is introduced for tensor form factors and it leads to a scale dependent term proportional to $C_F\alpha^2_s\ln(\mu^2)$ which can not be cancelled.
Thus it is reasonable that the scale dependence of tensor form factor $f_T$ at NLO is still large.

\begin{figure}[thb]          
\centering
\includegraphics[width=0.45\textwidth]{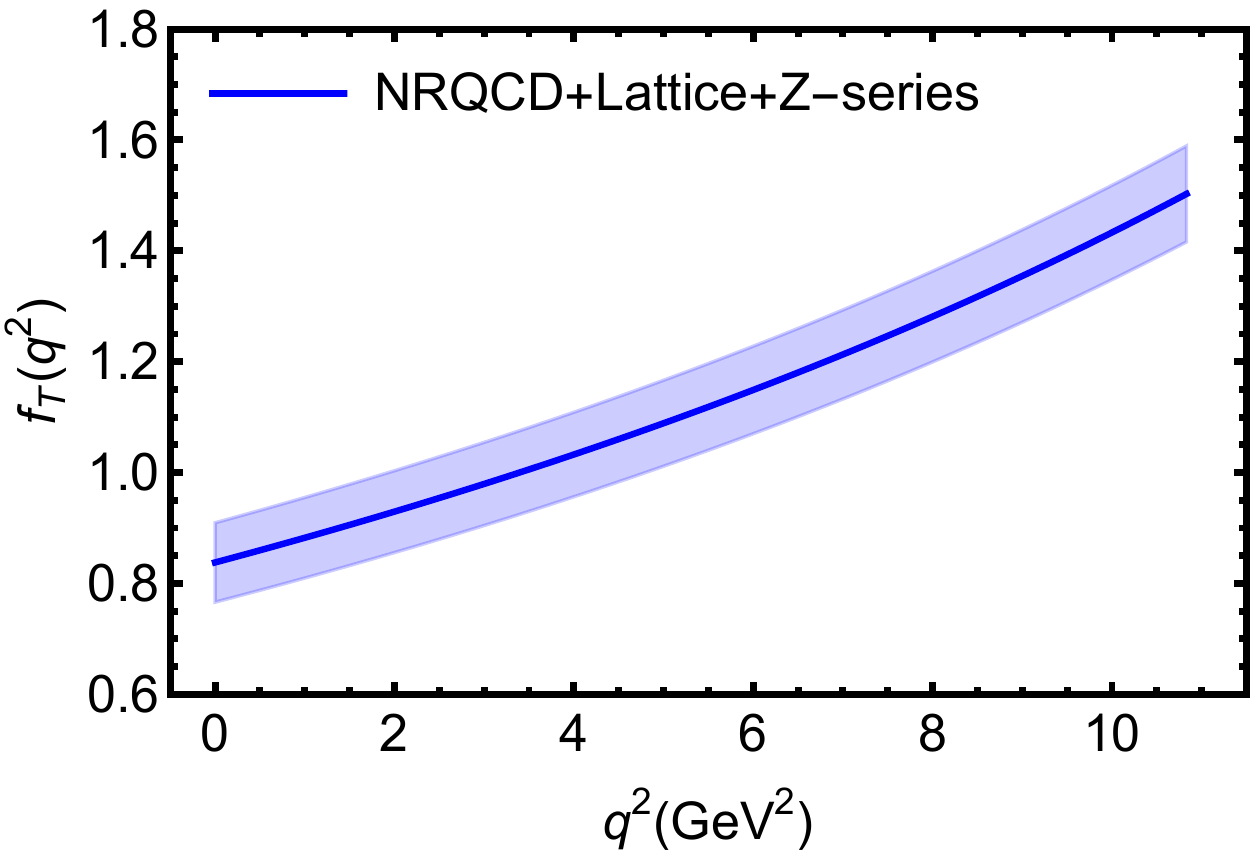}
\caption{The full curve of physical tensor form factor $f_T(q^2)$ for $B_c\to \eta_c$  transition with $0\leq q^2\leq (m_{B_c}- m_{\eta_c})^2$. The blue curve with
error band is the result of our polynomial fit in z-series combined with NRQCD calculation and the HPQCD lattice data of vector form factors for $B_c\to \eta_c$~\cite{Colquhoun:2016osw}. }
\label{fig:dependfTq2}
\end{figure}

\begin{figure}[thb]  
\centering
\includegraphics[width=0.45\textwidth]{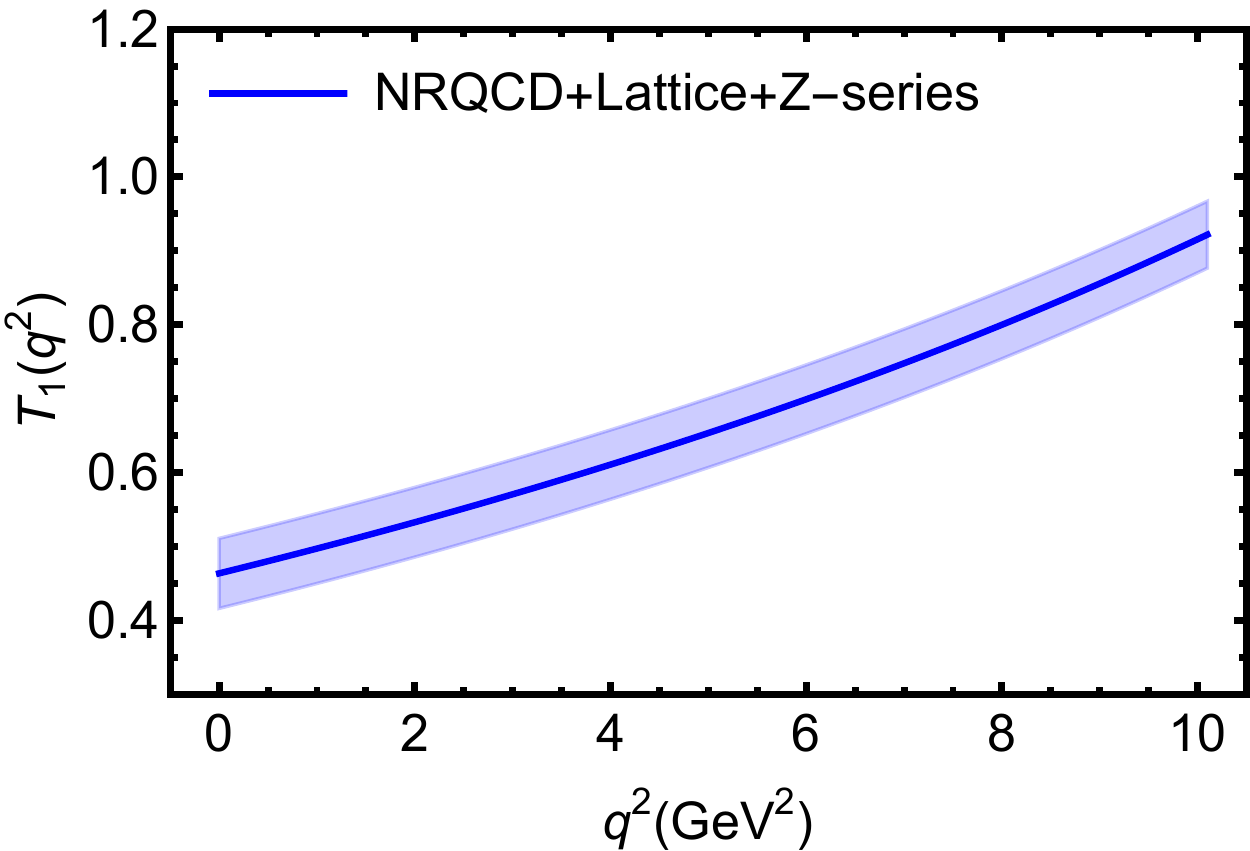}
\caption{The full curve of physical tensor form factor $T_1(q^2)$ for $B_c\to J/\psi$  transition with $0\leq q^2\leq (m_{B_c}- m_{J/\psi})^2$. The blue curve with
error band is the result of our polynomial fit in z-series combined with NRQCD calculation and the HPQCD lattice data of vector and axial-vector form factors for $B_c\to J/\psi$~\cite{Harrison:2020gvo}. }
\label{fig:dependT1q2}
\end{figure}

\begin{figure}[thb]  
\centering
\includegraphics[width=0.45\textwidth]{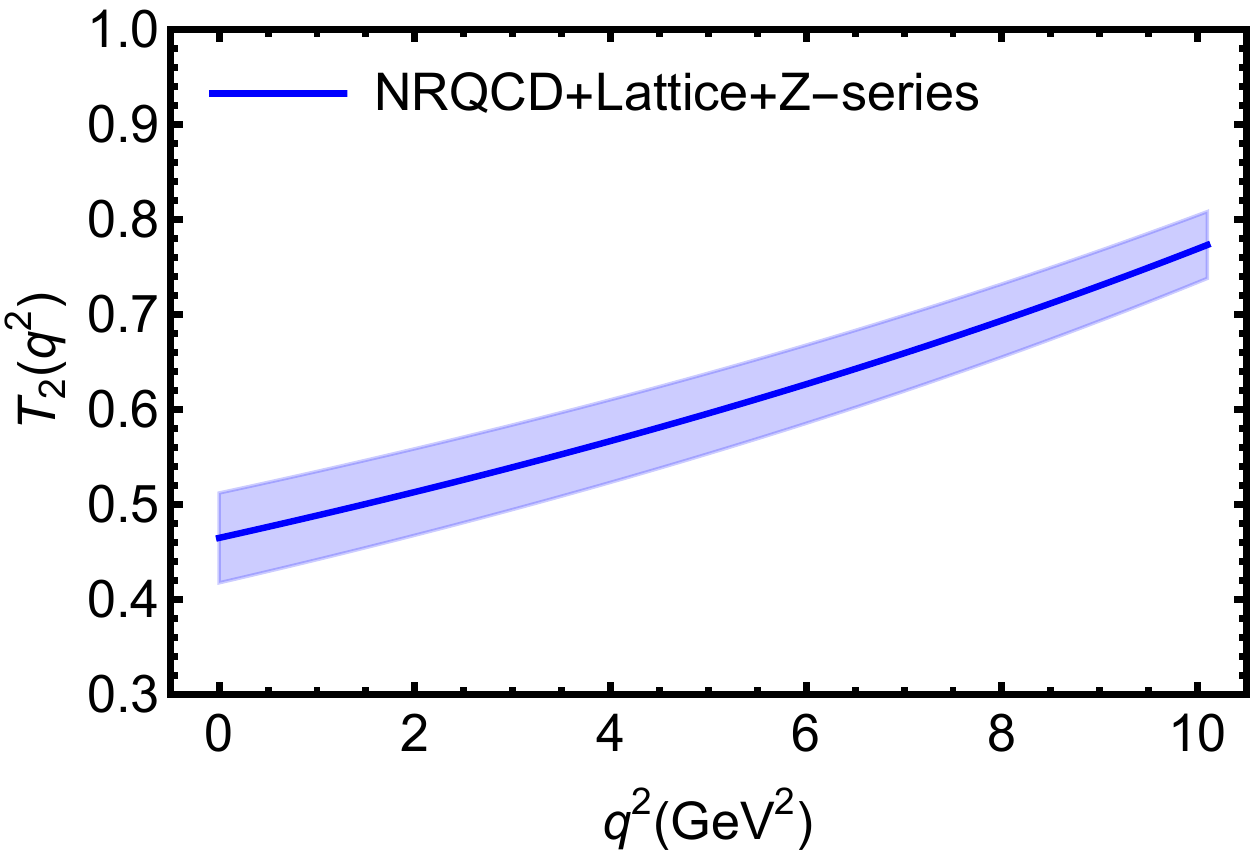}
\caption{The same as Fig.~\ref{fig:dependT1q2}, but for the physical tensor form factor $T_2(q^2)$ for $B_c\to J/\psi$. }
\label{fig:dependT2q2}
\end{figure}

\begin{figure}[thb]   
\centering
\includegraphics[width=0.45\textwidth]{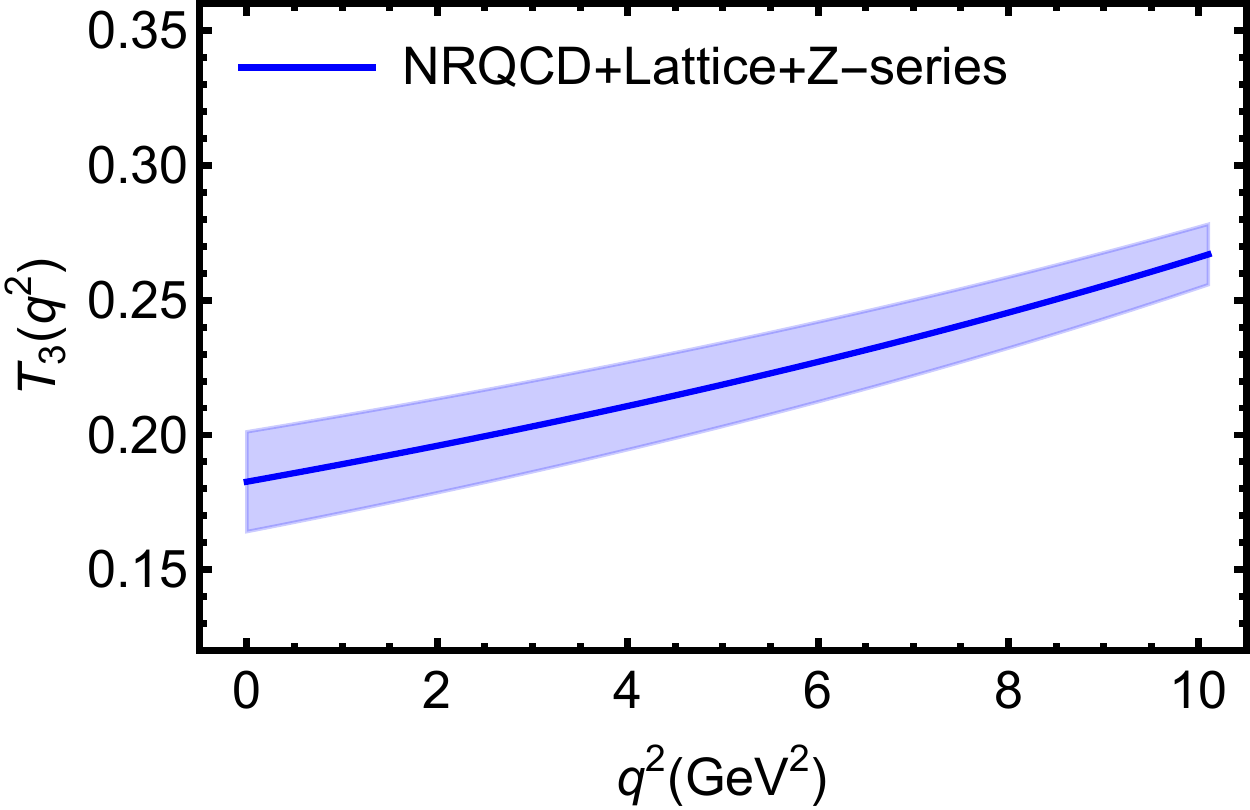}
\caption{The same as Fig.~\ref{fig:dependT1q2}, but for the physical tensor form factor $T_3(q^2)$ for $B_c\to J/\psi$. }
\label{fig:dependT3q2}
\end{figure}

Next we will focus on the theoretical predictions of tensor form factors in $B_c$ to a S-wave charmonium. To avoid the uncertainties from NRQCD LDMEs, we can employ the HPQCD lattice data of vector and axial-vector form factors in
$B_c$ to a S-wave charmonium~\cite{Colquhoun:2016osw,Harrison:2020gvo}. Combined the analytical expressions of vector, axial-vector and tensor factors in NRQCD framework, we can furthermore
obtain the tensor form factors.

However, the perturbative calculation in NRQCD is valid when the transferred momentum is large. Thus the analytical expressions of vector, axial-vector and tensor factors in NRQCD framework are not applicable for minimum momentum recoil region.  Thus we will use the Z-series method~\cite{Boyd:1997kz,Caprini:1997mu,Bourrely:2008za,Bharucha:2010im} to do the extrapolation. The tensor form factors can be rewritten as~\cite{Leljak:2019eyw,Hu:2019qcn}
\begin{eqnarray}
F_{i}(t) &=&\frac{1}{1-t / m_{R}^{2}} \sum_{k=0}^{\infty} \alpha_{k}^{i} z^{k}\left(t, t_{0}\right),
\end{eqnarray}
with
\begin{eqnarray}
z &=&\frac{\sqrt{t_{+}-t}-\sqrt{t_{+}-t_{0}}}{\sqrt{t_{+}-t}+\sqrt{t_{+}-t_{0}}},\\
t_{0} &=&t_{+}\left(1-\sqrt{1-\frac{t_{-}}{t_{+}}}\right), \\
t_{\pm} &=&\left(m_{B_{c}} \pm m_{\eta_c(J/\psi)}\right)^{2},
\end{eqnarray}
where  $t=q^{2}$. $m_{R}$ are the masses of the low-laying $B_{c}$ resonance. Here the series of parameter $z$
can be truncated to 2nd order because $z(q^2)\sim 0.02$ in $B_c$ to a S-wave charmonium~\cite{Wang:2018duy}.

We plot the  full curve of physical tensor form factors $f_T$ and $T_{1,2,3}$ for $B_c$ to a S-wave charmonium in Figs.~\ref{fig:dependfTq2}, \ref{fig:dependT1q2}, \ref{fig:dependT2q2}, and \ref{fig:dependT3q2}.
Our results  of  $B_c\to \eta_c$ and $B_c\to J/\psi$ tensor form factors  at maximum recoil $q^2=0$ are listed in Tab.~\ref{tab:FFs0}, together with the results from other literatures.
The uncertainties of our numerical results in Tab.~\ref{tab:FFs0} are from the HPQCD lattice data uncertainties of vector and axial-vector form factors~\cite{Colquhoun:2016osw,Harrison:2020gvo}.

\begin{table}[thb]
\begin{center}
\caption{Tensor form factors at maximum momentum recoil point $q^2=0$ calculated in this paper and other literatures.  }
\label{tab:FFs0}
\begin{tabular}{c c c c} \hline\hline
                               & $  f_{T}(0)$ &                $ T_{1}(0)= T_{2}(0)$ &                         $  T_{3}(0)$  \\\hline
NRQCD+Lattice                  & $0.85\pm{0.07}$               &  $0.46\pm{0.05}$                               &$0.18\pm{0.02}$ \; \\ \hline
CCQM\cite{Tran:2018kuv}        & $0.93$                        &  $0.56$                                         &$0.20$\; \\
CLFQM(type-II)\cite{Chang:2020wvs} & $0.90^{+0.17}_{-0.22}$    &  $0.56^{+0.16}_{-0.17}$                        &$0.19^{+0.03}_{-0.03}$ \; \\
QCDSR\cite{Leljak:2019eyw}      & $0.93\pm{0.07}$              &  $0.47\pm{0.04}$                               &$0.19\pm{0.01}$ \;  \\ \hline \hline
\end{tabular}
\end{center}
\end{table}

\section{ Conclusion }

While the lattice QCD have performed a state-of-the-art work on the vector and axial-vector form factors for  $B_c$ meson into a S-wave charmonium,
analyzing the pattern of  new physics  in $R(\eta_c)$ and $R(J/\psi)$ require more theoretical inputs. In this paper, we calculated the analytical
NLO corrections to tensor form factors for the transitions of $B_c$ meson into a S-wave charmonium, the $\eta_c$ and $J/\psi$.
The compact asymptotic expression of tensor form factors in heavy bottom
quark limit are presented. Combined the strict NLO results for vector, axial-vector, and tensor form factors and the HPQCD lattice data of vector and
axial-vector form factors, we obtained the full curve of the physical tensor form factors $f_T(q^2)$ and  $T_{1,2,3}(q^2)$
for  the considered $B_c\to \eta_c, J/\psi$  charmonia.  These results are useful to precisely study the semileptonic decays of $B_c$ meson into a S-wave charmonium such as
the $R(J/\psi)$ anomaly.

\section*{Acknowledgements}
This work is supported by NSFC under Grant No.~11775117 and No.~12075124,  and by Natural Science Foundation of Jiangsu under Grant No.~BK20211267.

\section*{Appendix}

\begin{appendix}

In this appendix, we have listed the analytical expression of tensor form factors  for  $B_c$ meson into a S-wave charmonium in the hierarchy heavy quark limit, i.e. i.e. $m_b\to\infty,~ m_c\to\infty, ~\mathrm{and }~z=m_c/m_b\to0$.
In general, we have $z=m_c/m_b$ and $s=1/(1-q^2/m_b^2)$. For $B_c\to \eta_c$ transition we have
\begin{widetext}
\begin{eqnarray}
\frac{f^\text{NLO}_{T}(z,s)}{f^\text{LO}_{T}(z,s)}
&=&1+\frac{\alpha _s}{4 \pi }\bigg\{ \left(\frac{11 C_A}{3}-\frac{2 n_f}{3}\right) \ln \frac{2 \mu ^2 s}{z m_b^2}-\frac{10 n_f}{9}-\frac{\ln z}{2}-\frac{\ln s}{2}-2 \ln 2+\frac{\pi ^2}{6}\nonumber\\
&&
+C_A \bigg[-\frac{\ln ^2 z}{4}+\left(-\frac{\ln s}{2}-\frac{3 \ln 2}{2}-\frac{1}{2}\right) \ln
   z+\left(\frac{1}{2}-2
   s\right) \text{Li}_2(1-2 s)+\left(s-\frac{1}{2}\right) \text{Li}_2(1-s)\nonumber\\
&&+\frac{1}{4} (-2 s-1) \ln^2 s+\left((-2 s-1) \ln
   2+\frac{s}{1-2 s}\right) \ln s+\left(-s-\frac{1}{2}\right) \ln^2 2+\frac{(1-3 s) \ln 2}{2 s-1}-\frac{1}{12} \pi ^2 (2 s+1)+\frac{67}{9}\bigg]\nonumber\\
&&
    +C_F \bigg[-\ln \frac{\mu
   ^2}{m_b^2}+\frac{5 \log ^2 z}{4}+\left(\frac{5 \ln s}{2}+6 \ln 2-\frac{23}{4}\right) \ln
   z+(4 s-1) \text{Li}_2(1-2 s)+(3-2 s) \text{Li}_2(1-s)\nonumber\\
&&+\left(s+\frac{9}{4}\right) \ln^2 s+\left((4 s+5) \ln 2+\frac{s (4 (34-11 s) s-101)+23}{4 (1-2 s)^2 (s-1)}\right) \ln s+\left(2 s+\frac{5}{2}\right) \ln
   ^2 2\nonumber\\
&&+\frac{(4 s (6 s-7)+7) \ln 2}{2 (1-2
   s)^2}+\frac{1}{12} \left(\pi ^2 (4 s+19)+\frac{6}{2 s-1}-207\right)\bigg]\bigg\}\,,
\end{eqnarray}
and at maximum recoil point ($s=1$ or $q^2=0$)
\begin{eqnarray}
\frac{f^\text{NLO}_{T}(z,1)}{f^\text{LO}_{T}(z,1)}
&=&1+
\frac{\alpha _s}{4 \pi } \bigg\{\left(\frac{11 C_A}{3}-\frac{2 n_f}{3}\right) \ln \frac{2 \mu ^2}{z m_b^2}-\frac{10 n_f}{9} -\frac{\ln z}{2}-2 \ln
   2+\frac{\pi ^2}{6}\nonumber\\
&&
+C_A \bigg[-\frac{1}{4} \ln
   ^2 z+\left(-\frac{3 \ln 2}{2}-\frac{1}{2}\right) \ln z
   -\frac{3 \ln^2 2}{2}-2 \ln 2-\frac{\pi ^2}{8}+\frac{67}{9}\bigg]\nonumber\\
&&
   +C_F
   \bigg[-\ln \frac{\mu ^2}{m_b^2}+\frac{5 \ln^2 z}{4}+\left(6 \ln 2-\frac{23}{4}\right) \ln z
   +\frac{9 \ln^2 2}{2}+\frac{3 \ln 2}{2}+\frac{5 \pi
   ^2}{3}-\frac{53}{4}\bigg]
  \bigg\}\,.
\end{eqnarray}
Note that the above  NLO result of $B_c\to \eta_c$ tensor form factor $f_T$ is in agreement with the previous calculation in Ref.~\cite{Bell:2006tz}.

For $B_c\to J/\psi$ transition we have

\begin{eqnarray}
\frac{T^\text{NLO}_{1}(z,s)}{T^\text{LO}_{1}(z,s)}
&=&1+\frac{\alpha _s}{4 \pi }\bigg\{
\left(\frac{11 C_A}{3}-\frac{2 n_f}{3}\right) \ln \frac{2 \mu ^2 s}{z m_b^2}-\frac{10 n_f}{9}\nonumber\\
&&
+C_A \bigg[-\frac{2 s \ln^2 z}{4 s+1}+\left(\left(\frac{1}{4 s+1}-1\right) \ln s+\left(\frac{6}{4 s+1}-2\right) \ln 2-\frac{6 s}{4
   s+1}\right) \ln z\nonumber\\
&&+\frac{(6 s-4)
   \text{Li}_2(1-2 s)}{4 s+1}+\left(\frac{5}{4 s+1}-1\right) \text{Li}_2(1-s)-\frac{s \ln^2 s}{4 s+1}+\left(-\frac{2 s \ln 2}{4 s+1}-\frac{6 s}{4 s+1}\right) \ln s\nonumber\\
&&-\frac{5 s \ln^2 2}{4 s+1}-\frac{6 s \ln
   2}{4 s+1}+\frac{268 s-3 \pi ^2 (3 s+2)+85}{36
   s+9}\bigg]\nonumber\\
&&
   +C_F \bigg[-\ln \frac{\mu ^2}{m_b^2}+\left(1-\frac{2}{4 s+1}\right) \ln^2 z+\left(\left(2-\frac{4}{4 s+1}\right) \ln s+\left(10-\frac{12}{4 s+1}\right) \ln
   2+\frac{1}{-4 s-1}-5\right) \ln z\nonumber\\
&&+\left(\frac{9}{4 s+1}-3\right) \text{Li}_2(1-2 s)+\left(4-\frac{8}{4
   s+1}\right) \text{Li}_2(1-s)+\left(\left(7-\frac{3}{4 s+1}\right) \ln 2-\frac{5}{4 s+1}-2\right) \ln s\nonumber\\
&&+\frac{6 s \ln^2 s}{4 s+1}+\frac{(22 s+2) \ln
   ^2 2}{4 s+1}+\left(9-\frac{8}{4 s+1}\right) \ln
   2+\frac{\pi ^2 (2 s-1)}{12 s+3}-17\bigg]\bigg\}\,,
\end{eqnarray}
\begin{eqnarray}
\frac{T^\text{NLO}_{1}(z,s)}{T^\text{LO}_{1}(z,s)}=\frac{T^\text{NLO}_{2}(z,s)}{T^\text{LO}_{2}(z,s)}=\frac{T^\text{NLO}_{3}(z,s)}{T^\text{LO}_{3}(z,s)}\,.
\end{eqnarray}
and at maximum recoil point ($s=1$ or $q^2=0$)
\begin{eqnarray}
\frac{T^\text{NLO}_{1}(z,1)}{T^\text{LO}_{1}(z,1)}
&=&1+
\frac{\alpha _s}{4 \pi } \bigg\{\left(\frac{11 C_A}{3}-\frac{2 n_f}{3}\right) \ln \frac{2 \mu ^2}{z m_b^2}-\frac{10 n_f}{9}\nonumber\\
&&
+C_A \bigg[-\frac{2}{5} \ln
   ^2 z+\left(-\frac{4 \ln 2}{5}-\frac{6}{5}\right) \ln z
   -\ln^2 2-\frac{6 \ln
   2}{5}+\frac{1}{90} \left(706-33 \pi ^2\right)\bigg]\nonumber\\
&&
   +C_F
   \bigg[-\ln \frac{\mu ^2}{m_b^2}+\frac{3 \ln^2 z}{5}+\left(\frac{38 \ln 2}{5}-\frac{26}{5}\right) \ln
   z
   +\frac{24 \ln^2 2}{5}+\frac{37 \ln 2}{5}+\frac{\pi ^2}{6}-17\bigg]
  \bigg\}\,,
\end{eqnarray}
\begin{eqnarray}
\frac{T^\text{NLO}_{1}(z,1)}{T^\text{LO}_{1}(z,1)}=\frac{T^\text{NLO}_{2}(z,1)}{T^\text{LO}_{2}(z,1)}=\frac{T^\text{NLO}_{3}(z,1)}{T^\text{LO}_{3}(z,1)}\,.
\end{eqnarray}

\end{widetext}
\end{appendix}


\end{document}